**Predicting New Iron Garnet Thin Films with Perpendicular Magnetic Anisotropy**

Saeedeh Mokarian Zanjani[1], Mehmet C. Onbaşlı[1,2,*]

[1] Graduate School of Materials Science and Engineering, Koç University, Sarıyer, 34450 Istanbul, Turkey.
[2] Department of Electrical and Electronics Engineering, Koç University, Sarıyer, 34450 Istanbul, Turkey.
* Corresponding Author: monbasli@ku.edu.tr

**Abstract:**

Perpendicular magnetic anisotropy (PMA) is a necessary condition for many spintronic applications like spin-orbit torques switching, logic and memory devices. An important class of magnetic insulators with low Gilbert damping at room temperature are iron garnets, which only have a few PMA types such as terbium and samarium iron garnet. More and stable PMA garnet options are necessary for researchers to be able to investigate new spintronic phenomena. In this study, we predict 20 new substrate/magnetic iron garnet film pairs with stable PMA at room temperature. The effective anisotropy energies of 10 different garnet films that are lattice-matched to 5 different commercially available garnet substrates have been calculated using shape, magnetoelastic and magnetocrystalline anisotropy terms. Strain type, tensile or compressive depending on substrate choice, as well as the sign and the magnitude of the magnetostriction constants of garnets determine if a garnet film may possess PMA. We show the conditions in which Samarium, Gadolinium, Terbium, Holmium, Dysprosium and Thulium garnets may possess PMA on the investigated garnet substrate types. Guidelines for obtaining garnet films with low damping are presented. New PMA garnet films with tunable saturation moment and field may improve spin-orbit torque memory and compensated magnonic thin film devices.

**Introduction**

With the development of sputtering and pulsed laser deposition of high-quality iron garnet thin films with ultralow Gilbert damping, researchers have been able to investigate a wide variety of magnetization switching and spin wave phenomena[1-3]. The key enabler in many of these studies has been Yttrium iron garnet ($Y_3Fe_5O_{12}$, YIG)[4] which has a very low Gilbert damping allowing spin waves to propagate over multiple millimeters across chip. YIG thin films are useful for spin wave device applications, but since their easy axes lie along film plane, their utility cannot be extended to different mechanisms such as spin-orbit torques, Rashba-Edelstein effect, logic devices, forward volume magnetostatic spin waves[1]. At the same time, to have reliable and fast response using low current densities as in spin-orbit torque switching, magnetization orientation needs to be perpendicular to the surface plane[5]. The possibility of having Dzyaloshinskii–Moriya interaction (DMI) in TmIG/GGG may enable stabilizing skyrmions and help drive skyrmion motion with pure spin currents[6].

There is a number of studies on tuning anisotropy or obtaining perpendicular magnetic anisotropy in insulator thin films[7-12]. Among the materials studied, insulating magnetic garnets whose magnetic properties can be tuned have been a matter of interest over the past decades[13-15] due to their low damping and high magnetooptical Faraday rotation. In order to obtain perpendicular magnetic anisotropy in magnetic garnets, one needs to engineer the anisotropy terms that give rise to out-of-plane easy axis. Angular dependence of total magnetization energy density is called magnetic anisotropy energy and consists of contributions from shape anisotropy, strain-induced (magnetoelastic) and magnetocrystalline anisotropy. A magnetic material preferentially relaxes its magnetization vector towards its easy axis, which is the least energy axis, when there is no external field bias. Such energy minimization process drives magnetic switching rates as well as the stability of total magnetization vector. Controlling magnetic anisotropy in thin film garnets not only offers researchers different testbeds for experimenting new PMA-based switching phenomena, but also allows the investigation of anisotropy-driven ultrafast dynamic magnetic response in thin films and nanostructures.

The most extensively studied garnet thin film is Yttrium Iron Garnet (YIG). YIG films display in-plane easy axis because of their large shape anisotropy and negligible magnetocrystalline anisotropy[3]. Although PMA of ultrathin epitaxial YIG films has been reported[16,17], the tolerance for fabrication condition variations for PMA YIG is very limited and strain effects were found to change magnetocrystalline anisotropy in YIG. Strain-controlled anisotropy has been observed in polycrystalline ultrathin YIG films[17,18]. In case of YIG thin film grown on Gadolinium Gallium Garnet (GGG), only partial anisotropy control has been possible through significant change in oxygen stoichiometry[19], which increases damping. Since the fabrication of high-quality and highly PMA YIG films is not easy for practical thicknesses on gadolinium gallium garnet substrates (GGG), researchers have explored tuning magnetic anisotropy by substituting Yttrium sites with other rare earth elements[20,21]. New garnet thin films that can exhibit PMA with different coercivities, saturation fields, compensation points and tunable Gilbert damping values must be developed in order to evaluate the effect of these parameters on optimized spintronic insulator devices.

Since the dominant anisotropy energy term is shape anisotropy in thin film YIG, some studies focus on reducing the shape anisotropy contribution by micro and nanopatterning[22-24]. Continuous YIG films were etched to form rectangular nanostrips with nanometer-scale thicknesses, as schematically shown on Fig. 1(a). Thus, least magnetic saturation field is needed along the longest dimension of YIG nanostrips. By growing ultrathin YIG, magnetoelastic strain contributions lead to a negative anisotropy field and thus PMA in YIG films[24]. As the length-to-thickness ratio decreases, the effect of shape anisotropy is reduced and in-plane easy axis is converted to PMA[17]. While reducing the effect of shape anisotropy is necessary, one also needs to use magnetoelastic anisotropy contribution to reorient magnetic easy axis towards out of film plane, as schematically shown on Fig. 1(b). Strain-induced perpendicular magnetic anisotropy in rare earth (RE) iron garnets, especially in YIG, has been demonstrated to overcome shape anisotropy[16,17,25,26]. If magnetoelastic anisotropy term induced by crystal lattice mismatch is larger than shape anisotropy and has opposite sign, then magnetoelastic anisotropy overcomes shape. Thus, the easy axis of the film becomes perpendicular to the film plane and the hysteresis loop becomes square-shaped with low saturation field[8]. One can also achieve PMA in other RE magnetic iron garnets due to their lattice parameter mismatch with their substrates. PMA has previously been achieved using Substituted Gadolinium Gallium Garnet (SGGG) as substrate and a Samarium Gallium Garnet (SmGG) ultrathin film as buffer layer under (and on) YIG[16]. In case of thicker YIG films (40nm), the magnetic easy axis becomes in-plane again. An important case shown by Kubota et.al[19] indicates that increasing in-plane strain ($\varepsilon_\parallel$) or anisotropy field ($H_a$) helps achieve perpendicular magnetic anisotropy. In ref.[8,19], they reported that if magnetostriction coefficient ($\lambda_{111}$) is negative and large enough to overcome shape anisotropy, and tensile strain is introduced to the thin film sample ($\varepsilon_\parallel > 0$), the easy axis becomes perpendicular to the sample plane as in the case of Thulium iron garnet ($Tm_3Fe_5O_{12}$, TmIG).

A different form of magnetoelastic anisotropy effect can be induced by using porosity in garnet thin films. Mesoporous Holmium Iron Garnet ($Ho_3Fe_5O_{12}$, HoIG) thin film on Si (001)[27] exhibits PMA due to reduced shape anisotropy, increased magnetostrictive and growth-induced anisotropy effects. Such combined effects lead to PMA in HoIG. In this porous thin film, the PMA was found to be independent of the substrate used, because the mechanical stress does not result from a lattice or thermal expansion mismatch between the substrate and the film. Instead, the pore-solid architecture itself imposes an intrinsic strain on the solution processed garnet film. This example indicates that the film structure can be engineered in addition to the substrate choices in order to overcome shape anisotropy in thin film iron garnets.

Another key method for controlling anisotropy is strain doping through substitutional elements and using their growth-induced anisotropy effects, as schematically shown on Fig. 1(c). Bi-doped yttrium iron garnet (Bi:YIG and Bi:GdIG) has been reported to possess perpendicular magnetic anisotropy due to the chemical composition change as the result of increased annealing temperature[21,28]. Another reason for PMA in these thin films is strain from GGG substrate[29]. Doping of oxides by Helium implantation was shown to reversibly and locally tune magnetic anisotropy[30]. For $Tb_xY_{3-x}Fe_5O_{12}$ (x=2.5, 2.0, 1.0, 0.37) samples grown by spontaneous nucleation technique[31], magnetic easy axis was found to change from [111] to [100] direction as Tb concentration was decreased. The first-order anisotropy constant $K_1$ undergoes a change of sign near 190K. Another temperature-dependent lattice distortion effect that caused anisotropy change was also reported for YIG films[32]. These results indicate that temperature also plays an important role in both magnetic compensation, lattice distortion and change in anisotropy.

In this study, we systematically calculate the anisotropy energies of 10 different types of lattice-matched iron garnet compounds epitaxially-grown as thin films ($X_3Fe_5O_{12}$, X = Y, Tm, Dy, Ho, Er, Yb, Tb, Gd, Sm, Eu) on commercially available (111)-oriented garnet substrates ($Gd_3Ga_5O_{12}$-GGG, $Y_3Al_5O_{12}$-YAG, $Gd_3Sc_2Ga_3O_{12}$-SGGG, $Tb_3Ga_5O_{12}$–TGG, $Nd_3Ga_5O_{12}$-NGG). Out of the 50 different film/substrate pairs, we found that 20 cases are candidates for room temperature PMA. Out of these 20 cases, 7 film/substrate pairs were experimentally tested and shown to exhibit characteristics originating from PMA. The remaining 13 pairs, to the best of our knowledge, have not been tested for PMA experimentally. We indicate through systematic anisotropy calculations that large strain-induced magnetic anisotropy terms may overcome shape when the films are highly strained. We use only the room temperature values of $\lambda_{111}$[33] and only report predictions for room temperature (300K). Throughout the rest of this study, the films are labelled as XIG (X = Y, Tm, Dy, Ho, Er, Yb, Tb, Gd, Sm, Eu), i.e. TbIG (Terbium iron garnet) or SmIG (Samarium iron garnet) etc. to distinguish them based on the rare earth element. Our model could accurately predict the magnetic easy axis in almost all experimentally tested garnet film/substrate cases provided that the actual film properties are entered in the model and that the experimental film properties satisfy cubic lattice matching condition to the substrate.

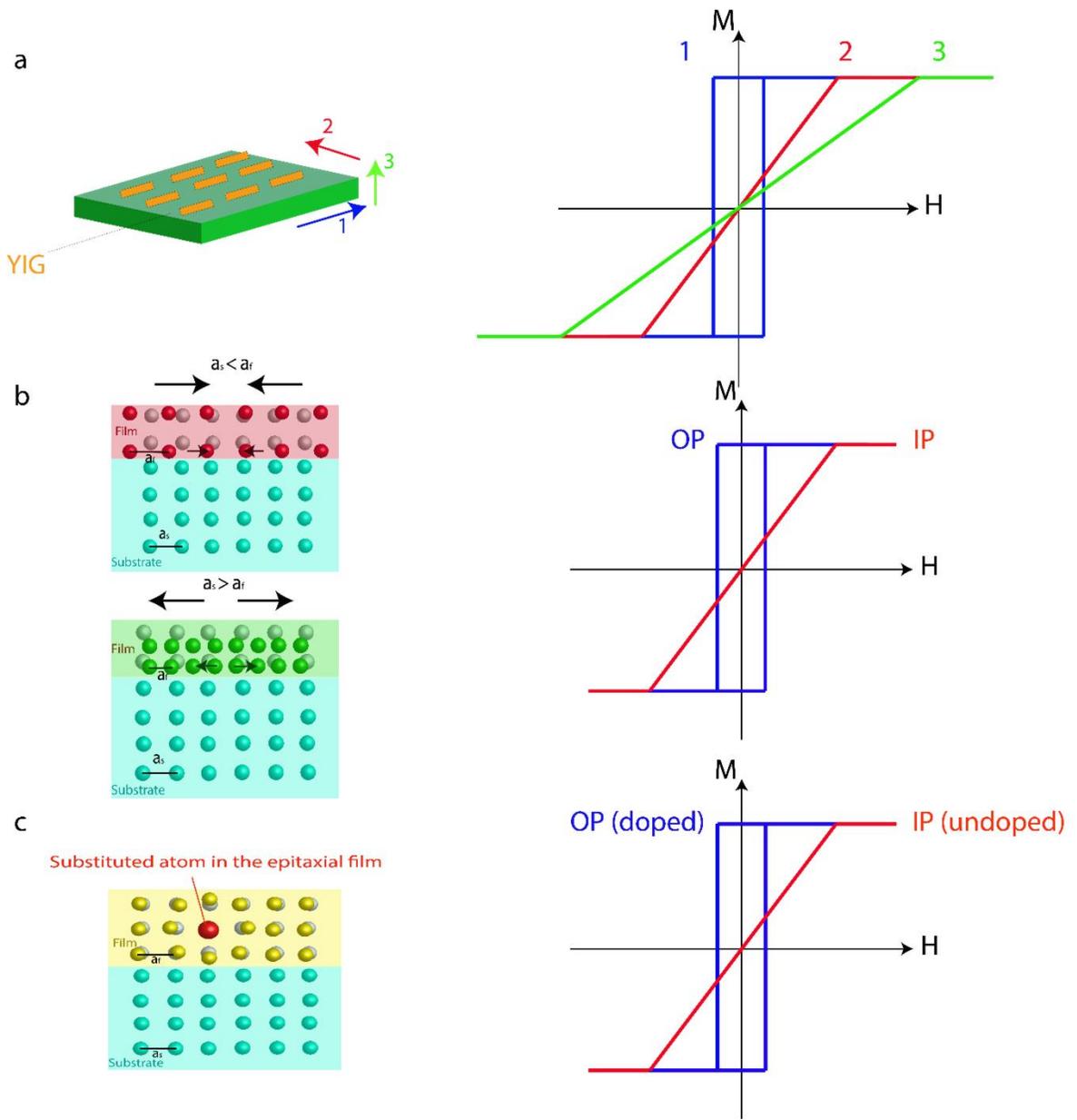

**Figure 1. Methods to achieve perpendicular magnetic anisotropy in iron garnet thin films.** (**a**) Micro/nano-patterning reduces shape anisotropy and magnetoelastic anisotropies overcome shape. (**b**) Large strain-induced anisotropy must overcome shape anisotropy to yield out-of-plane easy axis. (**c**) Substitutional doping in garnets overcome shape anisotropy by enhancing magnetocrystalline, growth-induced or magnetoelastic terms.

**Anisotropy energy density calculations**

Total anisotropy energy density contains three main contributions; according the Equation 1, shape anisotropy ($K_{shape}$), first order cubic magnetocrystalline anisotropy ($K_1$), and strain-induced (magnetoelastic) anisotropy ($K_{indu}$) parameters determine the total effective anisotropy energy density[16].

$$K_{eff} = K_{indu} + K_{shape} + K_1 \qquad (1)$$

In case of garnet film magnetized along [111] direction (i.e. on a 111 substrate), the magneto-elastic anisotropy energy density, resulting from magneto-elastic coupling is calculated by Equation 2:

$$K_{indu} = -\frac{3}{2}\lambda_{111}\sigma_{||} \qquad (2)$$

where $\lambda_{111}$ is magnetostriction constant along [111] direction and it is usually negative at room temperature[34]. In Eqn. 2, $\sigma_{||}$ is the in-plane stress induced in the material as a result of lattice mismatch between film and the substrate, and the in-plane stress is calculated from Equation 3[35]:

$$\sigma_{||} = \frac{Y}{1-\nu}\varepsilon_{||} \qquad (3)$$

where Y is elastic modulus, and $\nu$ is Poisson's ratio[36].

For calculation of in-plane strain, lattice parameter values obtained from the XRD characterization of the thin films are used. Equation 4 shows the strain relation as the lattice constant difference between the bulk form of the film and that of the substrate divided by the lattice constant for the bulk form of the film[16].

$$\varepsilon_{||} = \frac{a_{film} - a_{bulk}}{a_{film}} \qquad (4)$$

Assuming the lattice parameter of the thin film matches with that of the substrate, the lattice constant of substrate can be used as the lattice constant of thin film for calculation of strain in Equation 5[37]:

$$\varepsilon_{||} = \frac{a_{sub} - a_{film}}{a_{film}} \qquad (5)$$

The lattice constants used for the films and substrates examined for this study are presented on Table 1.

Shape anisotropy energy density depends on the geometry and the intrinsic saturation magnetic moment of the iron garnet material. Shape anisotropy has a demagnetizing effect on the total anisotropy energy density. These significant anisotropy effects can be observed in magnetic hysteresis loops and FMR measurements[24].

The most common anisotropies in magnetic materials are shape anisotropy and magneto-crystalline anisotropy[38,39]. Considering that the film is continuous, the shape anisotropy is calculated as[16]

$$K_{shape} = 2\pi M_s^2 \tag{6}$$

By obtaining the values of $M_s$ for rare earth iron garnets as a function of temperature[40,41], the value for shape anisotropy energy density have been calculated using Equation 6.

Intrinsic magnetic anisotropy[42], so called magnetocrystalline anisotropy, has the weakest contribution to anisotropy energy density compared to shape, and strain-induced anisotropies[9,11,16,19]. The values for the first order magnetocrystalline anisotropy is calculated and reported previously for rare earth iron garnets at different temperatures[43]. A key consideration in magnetic thin films is saturation field. In anisotropic magnetic thin films, the anisotropy fields have also been calculated using equation 7 as a measure of how much field the films need for magnetic saturation along the easy axis:

$$H_A = 2\frac{K_{eff}}{M_s} \tag{7}$$

**Table 1.** List of magnetic iron garnet thin films and garnet substrates available off-the-shelf used for this study. The fourth column shows the lattice constants used for calculating the magnetoelastic anisotropy values of epitaxial garnets on the given substrates.

| Garnet material | Chemical formula | Purpose | Bulk lattice constant (Å) |
|---|---|---|---|
| GGG | $Gd_3Ga_5O_{12}$ | Substrate | 12.383 |
| YAG | $Y_3Al_5O_{12}$ | Substrate | 12.005 |
| SGGG | $Gd_3Sc_2Ga_3O_{12}$ | Substrate | 12.480 |
| TGG | $Tb_3Ga_5O_{12}$ | Substrate | 12.355 |
| NGG | $Nd_3Ga_5O_{12}$ | Substrate | 12.520 |
| YIG | $Y_3Fe_5O_{12}$ | Film | 12.376 |
| TmIG | $Tm_3Fe_5O_{12}$ | Film | 12.324 |
| DyIG | $Dy_3Fe_5O_{12}$ | Film | 12.440 |
| HoIG | $Ho_3Fe_5O_{12}$ | Film | 12.400 |
| ErIG | $Er_3Fe_5O_{12}$ | Film | 12.350 |
| YbIG | $Yb_3Fe_5O_{12}$ | Film | 12.300 |

| | | | | |
|---|---|---|---|---|
| TbIG | Tb$_3$Fe$_5$O$_{12}$ | Film | 12.460 |
| GdIG | Gd$_3$Fe$_5$O$_{12}$ | Film | 12.480 |
| SmIG | Sm$_3$Fe$_5$O$_{12}$ | Film | 12.530 |
| EuIG | Eu$_3$Fe$_5$O$_{12}$ | Film | 12.500 |

**Results and Discussion**

Tables 1 and 2 list in detail the parameters used and the calculated anisotropy energy density terms for magnetic rare earth iron garnets at 300K. These tables show only the cases predicted to be PMA out of a total of 50 film/substrate pairs investigated. The extended version of Tables 1 and 2 for all calculated anisotropy energy density terms for all combinations of the 50 film/substrate pairs are provided in the supplementary tables. The tabulated values for saturation magnetization[40,41] and lattice parameters[44] have been used for the calculations. In this study, we assumed the value of Young's modulus and Poisson ratio as $2.00\times10^{12}$ dyne·cm$^{-2}$ and 0.29 for all garnet types, respectively, based on ref.[36]. We also assume that the saturation magnetization, used for calculation of shape anisotropy, does not change with the film thickness. The saturation magnetization ($M_s$) values and shape anisotropy for iron garnet films are presented in the third and fourth columns, respectively. The stress values for fully lattice-matched films σ calculated using equation 3 and magnetostriction constants of the films, $\lambda_{111}$, are presented on columns 6 and 7. Magnetoelastic anisotropy $K_{indu}$, magnetocrystalline anisotropy energy density $K_1$, and the total magnetic anisotropy energy density $K_{eff}$ are calculated and listed on columns 8, 9 and 10, respectively. $H_A$ on column 11 is the anisotropy field (the fields required to saturate the films).

**Table 2.** Anisotropy energy density parameters calculation results. Rare earth iron garnets on GGG ($a_s$=12.383Å), YAG (Y$_3$Al$_5$O$_{12}$, $a_s$=12.005Å), SGGG ($a_s$=12.48Å) and TGG (Tb$_3$Ga$_5$O$_{12}$, $a_s$=12.355Å), and NGG (Nd$_3$Ga$_5$O$_{12}$, $a_s$=12.509Å) substrates, with $K_{eff}$ < 0, are presented.

| Film | Substrate | $M_s$ (emu·cm$^{-3}$) | $K_{shape}$ (erg·cm$^{-3}$) ($\times 10^3$) | ε ($\times 10^{-3}$) | σ (dyn·cm$^{-2}$) ($\times 10^{10}$) | $\lambda_{111}$ ($\times 10^{-6}$) | $K_{indu}$ (erg·cm$^{-3}$) ($\times 10^4$) | $K_1$ (300K) (erg·cm$^{-3}$) ($\times 10^3$) | $K_{eff}$ (erg·cm$^{-3}$) ($\times 10^3$) | $H_A$ (Oe) ($\times 10^3$) |
|---|---|---|---|---|---|---|---|---|---|---|
| DyIG | GGG | 31.85 | 6.37 | -4.58 | -1.29 | -5.9 | -11.4 | -5.00 | -113 | -7.09 |
| HoIG | GGG | 55.73 | 19.5 | -1.37 | -0.386 | -4 | -2.3 | -5.00 | -8.66 | -0.311 |
| GdIG | GGG | 7.962 | 0.398 | -7.77 | -2.19 | -3.1 | -10.2 | -4.10 | -106 | -26.5 |
| SmIG | GGG | 140 | 123 | -11.7 | -3.30 | -8.6 | -42.6 | -17.4 | -321 | -4.58 |
| YIG | YAG | 141.7 | 126 | -30.0 | -8.44 | -2.4 | -30.4 | -6.10 | -184 | -2.60 |

| | | | | | | | | | |
|---|---|---|---|---|---|---|---|---|---|
| TmIG | YAG | 110.9 | 77.2 | -25.9 | -7.29 | -5.2 | -56.9 | -5.80 | -497 | -8.97 |
| DyIG | YAG | 31.85 | 6.37 | -35.0 | -9.85 | -5.9 | -87.2 | -5.00 | -870 | -54.7 |
| HoIG | YAG | 55.73 | 19.5 | -31.9 | -8.97 | -4 | -53.8 | -5.00 | -524 | -18.8 |
| ErIG | YAG | 79.62 | 39.8 | -27.9 | -7.87 | -4.9 | -57.8 | -6.00 | -545 | -13.7 |
| YbIG | YAG | 127.3 | 102 | -24.0 | -6.76 | -4.5 | -45.6 | -6.10 | -360 | -5.66 |
| GdIG | YAG | 7.962 | 0.398 | -38.1 | -10.7 | -3.1 | -49.9 | -4.10 | -502 | -126 |
| SmIG | YAG | 140 | 123 | -41.9 | -11.8 | -8.6 | -152.3 | -17.4 | -1420 | -20.2 |
| TbIG | SGGG | 15.92 | 1.59 | 1.61 | 0.452 | 12 | -8.14 | -8.20 | -88.0 | -11.1 |
| GdIG | SGGG | 7.962 | 0.398 | 0.00 | 0.00 | -3.1 | 0.00 | -4.10 | -3.70 | -0.930 |
| SmIG | SGGG | 140 | 123 | -3.99 | -1.12 | -8.6 | -14.5 | -17.4 | -39.3 | -0.562 |
| DyIG | TGG | 31.85 | 6.37 | -6.83 | -1.92 | -5.9 | -17.0 | -5.00 | -169 | -10.6 |
| HoIG | TGG | 55.73 | 19.5 | -3.63 | -1.02 | -4 | -6.13 | -5.00 | -46.8 | -1.68 |
| GdIG | TGG | 7.962 | 0.398 | -10.0 | -2.82 | -3.1 | -13.1 | -4.10 | -135 | -33.9 |
| SmIG | TGG | 140 | 123 | -14.0 | -3.93 | -8.6 | -50.8 | -17.4 | -402 | -5.74 |
| TbIG | NGG | 15.92 | 1.59 | 3.93 | 1.11 | 12 | -19.9 | -8.20 | -206 | -25.9 |

In this study, we take the same sign convention as in ref. 16 and the films exhibit PMA when $K_{eff}$ < 0. So for obtaining PMA, negative and large values for anisotropy energy density are desired. As all the garnets (except TbIG) possess negative magnetostriction constants at room temperature, the sign of the strain (tensile or compressive) determines whether the induced anisotropy is negative or positive. In the literature[16,20,45,46] however, we observe that PMA was defined for either positive or negative effective anisotropy energy density ($K_{eff}$). This inconsistency may cause confusion among researchers. Thermodynamically, a higher energy means an unstable state with respect to lower energy cases. Easy axis, by definition, is the axis along which the magnetic material can be saturated with lowest external field or lowest total energy. A magnetic material would thus spontaneously minimize its energy and reorient its magnetic moment along the easy axis. As a result, we use here $K_{eff}$ < 0 for out-of-plane easy axis. Due to the thermodynamic arguments mentioned above, we suggest researchers to use $K_{eff}$ < 0 definition for PMA.

**Effect of Substrate on Anisotropy Energy Density**

Changing the substrate alters the strain in the film, which also changes strain-induced anisotropy in the film. Figure 2 shows the calculated anisotropy energy density of rare earth iron garnet thin films grown on five commercially available different substrates: Gadolinium Gallium Garnet ($Gd_3Ga_5O_{12}$, GGG), Yttrium Aluminum Garnet ($Y_3Al_5O_{12}$, YAG), Substituted Gadolinium

Gallium Garnet ($Gd_3Sc_2Ga_3O_{12}$, SGGG), Terbium Gallium Garnet ($Tb_3Ga_5O_{12}$, TGG), and Neodymium Gallium Garnet ($Nd_3Ga_5O_{12}$, NGG). As shown on Fig. 2(a), when grown on GGG substrate; Dysprosium Iron Garnet (DyIG), Holmium Iron Garnet (HoIG), Gadolinium Iron Garnet (GdIG), and Samarium Iron Garnet (SmIG) possess compressive strain ($a_{film} > a_{substrate}$). Considering the large negative magnetostriction constant ($\lambda_{111}$) for each case, the strain-induced anisotropy energy densities are estimated to cause negative total effective anisotropy energy density. As a result, DyIG, HoIG, GdIG, and SmIG on GGG are predicted to be PMA cases.

Based on the shape, magnetoelastic and magnetocrystalline anisotropy terms (room temperature $K_1$), Thulium iron garnet (TmIG) on GGG (111) is estimated to be in-plane easy axis although unambiguous experimental evidence indicates that TmIG grows with PMA on GGG (111) [1,19]. The fact that only considering shape, magnetocrystalline and magnetoelastic anisotropy terms does not verify this experimental result suggests that the PMA in TmIG/GGG (111) may originate from a different anisotropy term such as surface anisotropy, growth-induced or stoichiometry-driven anisotropy. Since the films used in the experiments are less than 10 nm or 5-8 unit cells thick, surface effects may become more significant and may require density functional theory-based predictions to account for surface anisotropy effects.

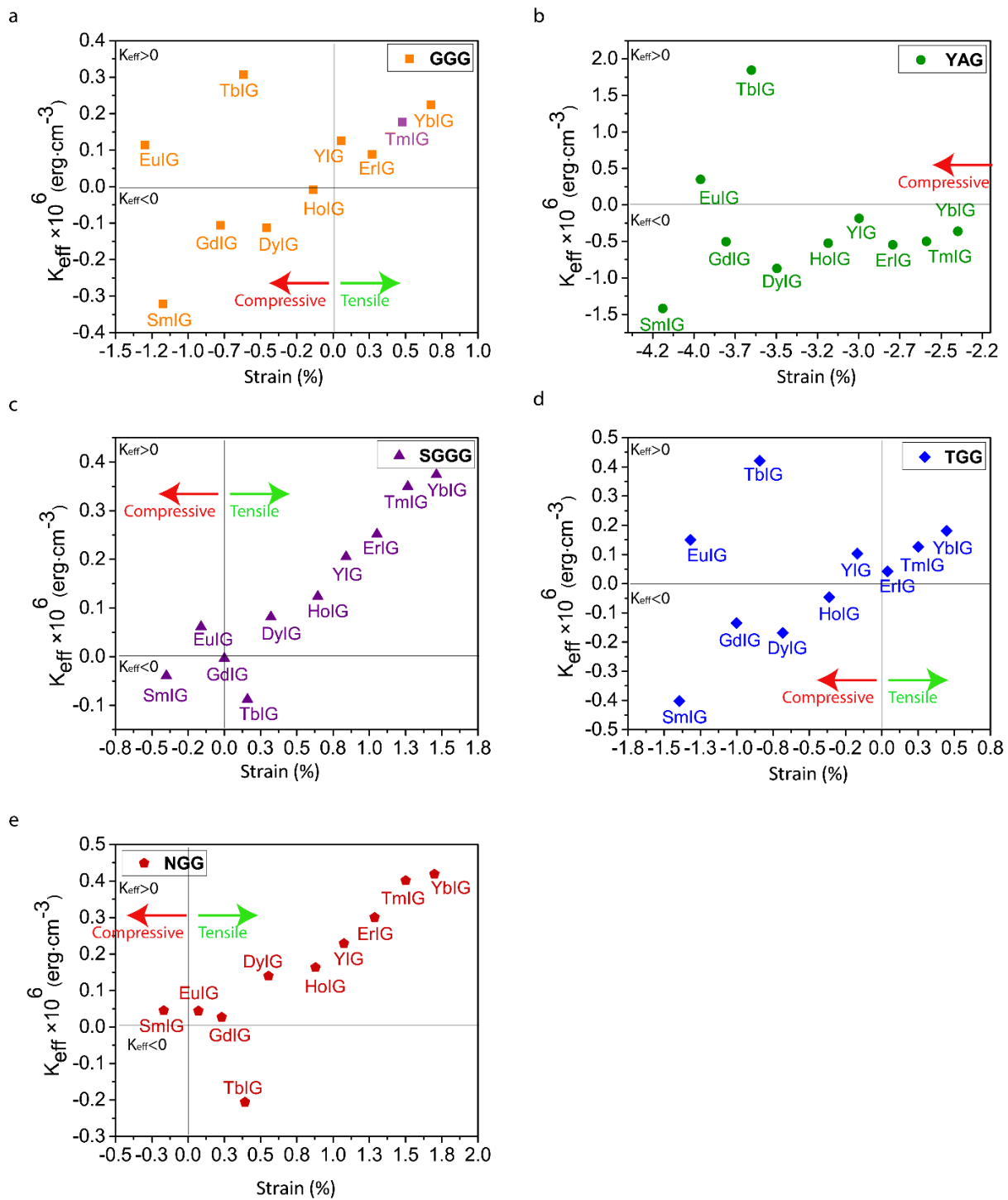

**Figure 2.** Calculated effective anisotropy values for each rare earth iron garnet thin film when they are epitaxially grown on (**a**) GGG, (**b**) YAG, (**c**) SGGG, (**d**) TGG, (**e**) NGG substrates. Note that the scales of the axes are different in each part.

Yttrium Aluminum Garnet (YAG) is a substrate with smaller lattice parameter than all the rare earth iron garnet films considered. With a substrate lattice parameter of $a_s$=12.005Å, YAG causes significant and varying amounts of strain on YIG ($a_f$=12.376Å), TmIG ($a_f$=12.324Å), DyIG ($a_f$=12.440Å), HoIG ($a_f$=12.400Å), ErIG ($a_f$=12.350Å), YbIG ($a_f$=12.300Å), TbIG ($a_f$=12.460Å), GdIG ($a_f$=12.480Å), SmIG ($a_f$=12.530Å) and EuIG ($a_f$=12.500Å). Strain from YAG substrate yields negative strain-induced anisotropy energy density for these films. The strain-induced anisotropy term overcomes the shape anisotropy in these garnets when they are grown on YAG. Consequently, effective anisotropy energy densities become negative and these garnet films are estimated to possess perpendicular magnetic anisotropy. In the exceptional cases of Terbium Iron Garnet (TbIG) and Europium Iron Garnet (EuIG), compressive strain is not enough to induce negative strain anisotropy because the magnetostriction coefficients of TbIG and EuIG are positive. So the strain-induced anisotropy terms are also positive for both TbIG ($K_{indu(TbIG)} = 1.85\times10^6$ erg·cm$^{-3}$) and EuIG ($K_{indu(EuIG)} = 3.01\times10^5$ erg·cm$^{-3}$) and do not yield PMA.

Other potential PMA garnets as a film on SGGG substrate are GdIG, TbIG, and SmIG. TbIG and GdIG cases are particularly interesting as growth conditions of these materials can be further optimized to achieve room temperature compensation and zero saturation magnetization. This property enables PMA garnet-based room temperature terahertz magnonics. The lattice parameters of GdIG ($a_f$=12.480Å) and SGGG ($a_s$=12.48Å) match exactly, so the in-plane strain value is zero and the effect of strain-induced anisotropy is eliminated completely. Consequently, due to small value for saturation magnetization of GdIG, shape anisotropy ($3.98 \times10^2$ erg·cm$^{-3}$) cannot compete with magnetocrystalline anisotropy ($-4.1\times10^3$ erg·cm$^{-3}$). In other words, in this case, the influence of magnetocrystalline anisotropy is not negligible compared to the other anisotropy terms. Consequently, the anisotropy energy density is negative for GdIG when grown on SGGG due to the influence of magnetocrystalline anisotropy energy density.

One other candidate for a PMA rare earth iron garnet on SGGG substrate is SmIG. Since the film lattice parameter is greater than that of the substrate, compressive strain ($-3.99 \times10^{-3}$) is induced in the film such that the resulting anisotropy energy density possesses a negative value of an order of magnitude ($-1.45\times10^5$ erg·cm$^{-3}$) comparable to the shape anisotropy energy density ($1.23\times10^5$ erg·cm$^{-3}$). With its relatively large magnetocrystalline anisotropy energy density ($-1.74\times10^4$ erg·cm$^{-3}$), SmIG has a perpendicular magnetic anisotropy due to negative value for effective anisotropy energy density (-4.80×10$^5$ erg·cm$^{-3}$). TbIG film on SGGG substrate is a PMA candidate with positive strain and this film was also recently experimentally demonstrated to have PMA[47].

Since TbIG's lattice constant is smaller than that of the substrate, the film becomes subject to tensile strain (+1.61 ×10$^{-3}$). Since TbIG also has a positive $\lambda_{111}$ (in contrast to that of SmIG), the film's magnetoelastic anisotropy term becomes large and negative and overcomes the shape anisotropy. In case of TbIG, magnetocrystalline anisotropy alone overcomes shape and renders the film PMA on SGGG. With the additional magnetoelastic anisotropy contribution (-8.14×10$^4$ erg·cm$^{-3}$), significant stability of PMA can be achieved.

Terbium Gallium Garnet (TGG) is a substrate with lattice parameter (a$_s$=12.355Å) such that it can induce tensile strain on TmIG, ErIG and YbIG and it induces compressive strain on the rest of the rare earth iron garnets (YIG, DyIG, HoIG, TbIG, GdIG, SmIG, EuIG). In none of the tensile-strained cases, PMA can be achieved since the sign of the magnetoelastic anisotropy is positive and has the same sign as the shape anisotropy. Among the compressively strained cases, YIG, TbIG and EuIG are found to have weak magnetoelastic anisotropy terms which cannot overcome shape. As a result, YIG, TbIG and EuIG on TGG substrate are expected to have in-plane easy axis. DyIG, HoIG, GdIG and SmIG films on TGG achieve large and negative effective total anisotropy energy densities due to their negative $\lambda_{111}$ values. In addition, since the materials have compressive strain, the signs cancel and lead to large magnetoelastic anisotropy energy terms that can overcome shape in these materials. So these cases are similar to the conditions explained for GGG substrate, on which only DyIG, HoIG, GdIG, and SmIG films with compressive strain can gain a large negative strain-induced anisotropy energy density which can overcome shape anisotropy.

Neodymium gallium garnet (NGG) is a substrate [45] used for growing garnet thin films by pulsed laser deposition method. NGG has large lattice constant compared with the rest of the bulk rare earth garnets and yield compressive strain in all rare earth garnets investigated except for Samarium iron garnet (SmIG). For all cases other than SmIG, the sign of the magnetoelastic anisotropy term is determined by the respective $\lambda_{111}$ for each rare earth iron garnet. YIG, TmIG, DyIG, HoIG, ErIG, YbIG cases have positive magnetoelastic anisotropy terms, which lead to easy axes along their film planes. For SmIG and EuIG on NGG, magnetoelastic strain and anisotropy terms are not large enough to overcome large shape anistropy. For GdIG, the weak tensile strain on NGG substrates actually causes in-plane easy axes as magnetoelastic strain offsets the negative magnetocrystalline anisotropy. The only rare earth iron garnet that can achieve PMA on NGG is Terbium Iron Garnet (TbIG) due to its large negative strain-induced anisotropy energy density ($-2.44\times10^5$ erg·cm$^{-3}$). Its large and negative magnetoelastic anisotropy can offset shape ($1.59\times10^3$ erg·cm$^{-3}$) and first order magnetocrystalline anisotropy term ($-8.20\times10^3$ erg·cm$^{-3}$), leading to a large negative effective magnetic anisotropy energy density ($-2.51\times10^5$ erg·cm$^{-3}$). Consequently, we predict that growing TbIG on NGG substrate may yield PMA.

Figure 3 shows the substrates on which one may expect PMA rare earth iron garnet films (or negative $K_{eff}$) due to strain only. Figures 3(a)-(d) compare the calculated effective energy densities as a function of strain type and sign for YIG, TmIG, YbIG, TbIG. For comparing the calculation results presented here with the experimentally reported values for the anisotropy energy density of TmIG, we added the $K_{eff}$ directly from the experimental data in[20] to Fig. 3(b). As shown on Fig. 3(b), the experimental TmIG thin film shows positive $K_{eff}$ as the result of tensile in-plane strain and large negative magnetostriction constant.

Figure 4(a)-(f) shows the calculated effective energy densities as a function of strain type and sign for GdIG, SmIG, EuIG, HoIG, DyIG, ErIG, respectively. $K_{eff}$ may get a positive or a negative value in both compressive and tensile strain cases due to varying signs of $\lambda_{111}$ constants of rare earth iron garnets. In almost all cases that yield PMA on the given substrates, PMA iron garnets form under compressive lattice strain. The only exceptions in which tensile strain can yield PMA in garnet thin films is TbIG on SGGG and TbIG on NGG. In both of those cases, a small tensile strain enhances PMA but the magnetocrystalline anisotropy could already overcome shape and yield PMA without lattice strain. Therefore, experimental studies should target compressive lattice strain.

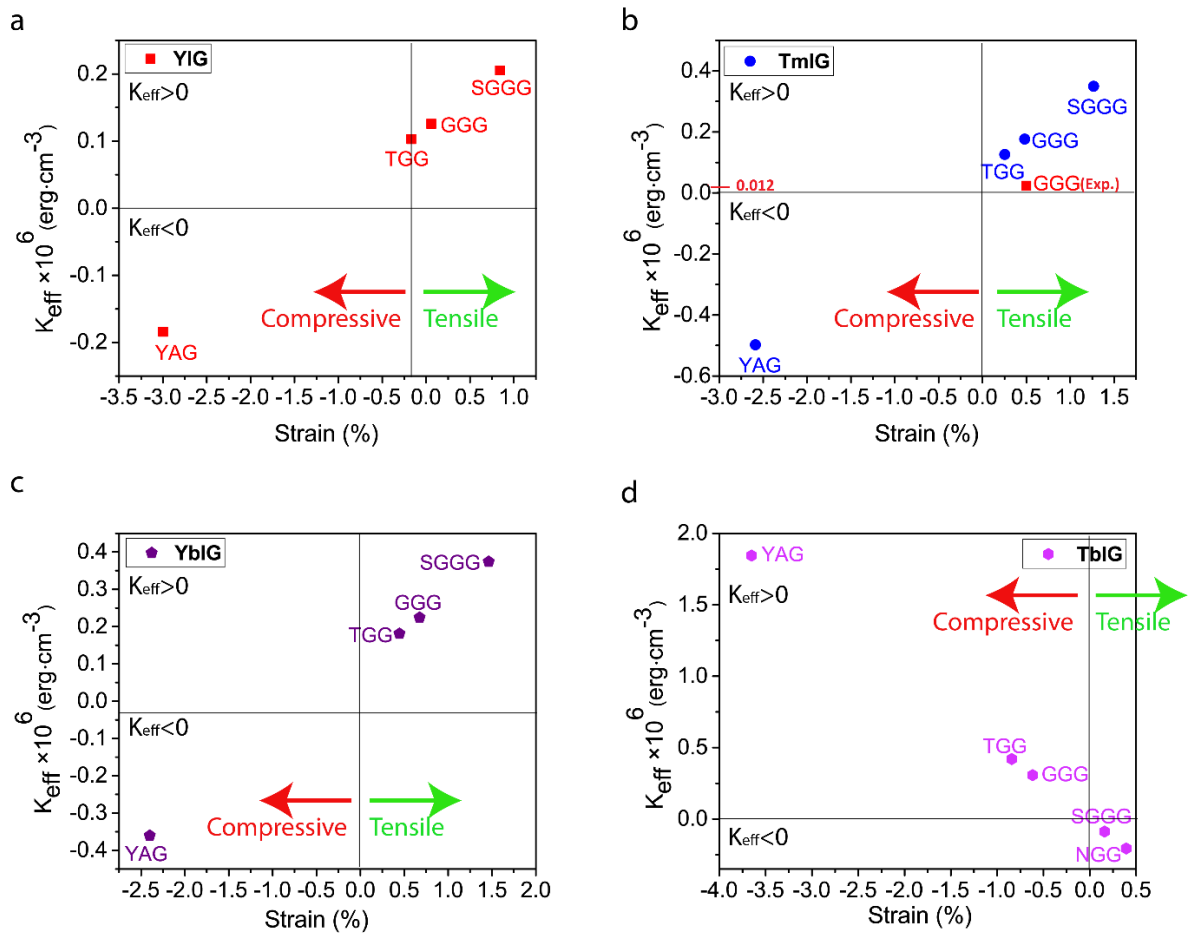

**Figure 3.** Effect of substrate strain on the effective anisotropy energy densities of (**a**) YIG, (**b**) TmIG, the data inserted on the graph, with red square symbol, GGG (Exp.), is the experimental value of effective anisotropy energy of TmIG on GGG based on ref.[20] (**c**) YbIG, (**d**) TbIG. Note that the axes scales are different in each part.

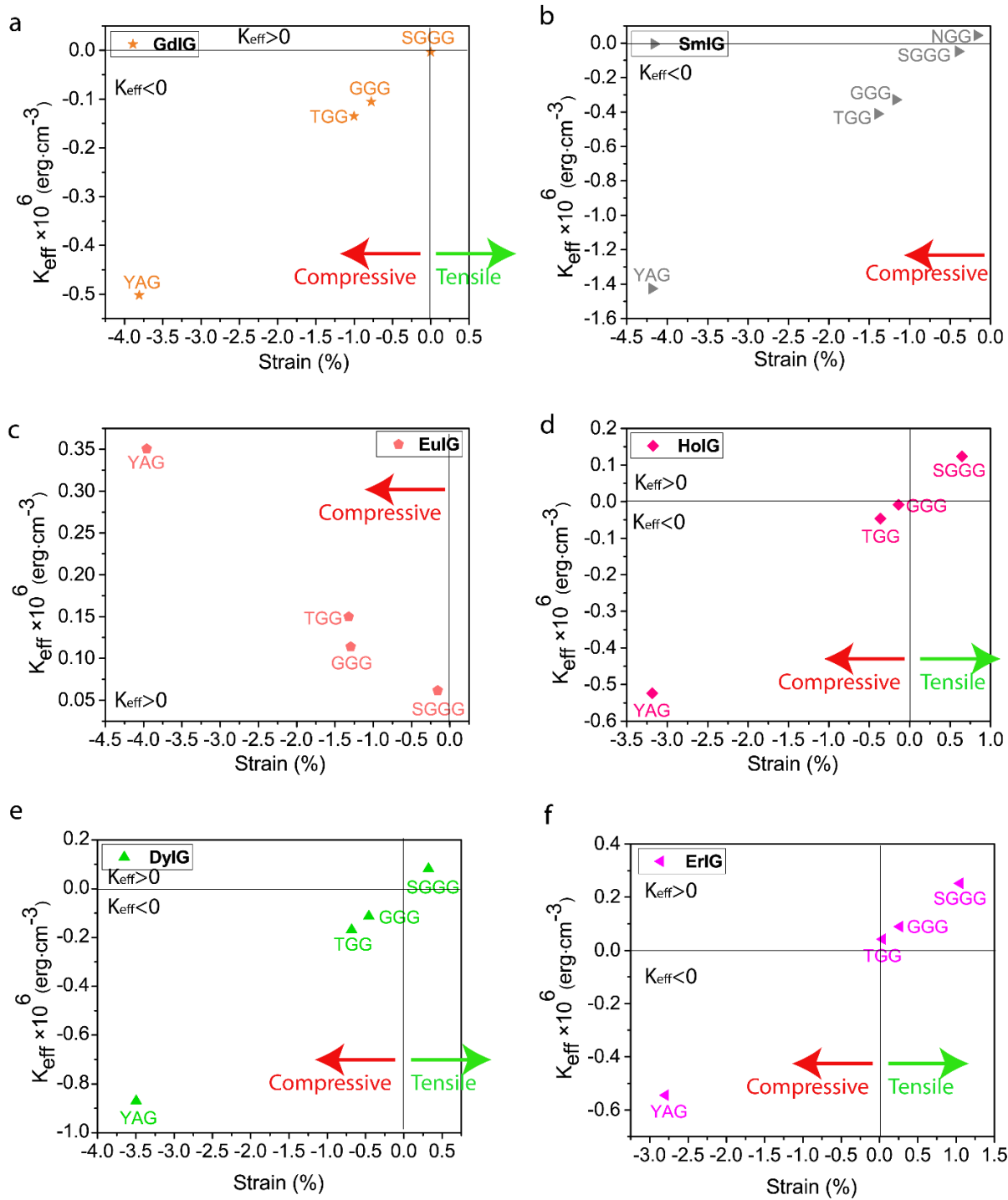

**Figure 4.** (**a**) GdIG, (**b**) SmIG, (**c**) EuIG, (**d**) HoIG, (**e**) DyIG, (**f**) ErIG films on GGG, YAG, SGGG, TGG, and NGG substrates. Note that the axes scales are different in each part.

Based on Fig. 2-4, the calculations in this paper numerically match with the reported values in the experimental demonstrations in literature both in sign and order of magnitude. However, since there are inconsistent sign conventions for predicting the magnetic anisotropy state of the iron garnet samples in the literature so far, some of the previous studies draw different conclusions on the anisotropy despite the similar $K_{eff}$.

In case of TmIG, as shown in Fig. 3(b), our model predicts that there is a tensile strain-induced anisotropy resulting from the difference in film and substrate lattice parameters and the film's negative magnetostriction constant. The experimental results of magnetic anisotropy in TmIG/GGG[8,20,48] are consistent with our model predictions. Previous studies identify PMA, if the film $K_{eff}$ is positive. A shortcoming of this approach is that such a definition would also identify YIG/GGG as PMA although its in-plane easy axis behavior has been experimentally demonstrated in numerous studies[3,32,49]. $K_{eff} < 0$ for PMA definition would be thermodynamically more appropriate and would also accurately explain almost all cases including YIG/GGG. Further explanation about TmIG exceptional case is included in the Supplementary Information.

**Sensitivity of Anisotropy to Variations in Saturation Magnetic Moment and Film Relaxation**

The films with predicted effective anisotropy energy and field may come out differently when fabricated due to unintentional variability in fabrication process conditions, film stoichiometry (rare earth ion to iron ratio and oxygen deficiency) as well as process-induced non-equilibrium phases in the garnet films. These changes may partially or completely relax the films or increase strain further due to secondary crystalline phases. Practical minor changes in strain may dramatically alter both the sign and the magnitude of magnetoelastic anisotropy energy and may cause a film predicted as PMA to come out with in-plane easy axis. On the other hand, off-stoichiometry may cause reduction in saturation magnetic moment. Reduction in saturation magnetic moment decreases shape anisotropy term quadratically ($K_{shape} = 2\pi M_s^2$), which implies that a 10% reduction in $M_s$ leads to a 19% decrease in $K_{shape}$ and anisotropy field may increase ($H_A = 2K_{eff}/M_s$). Therefore, sample fabrication issues and the consequent changes in anisotropy terms may weaken or completely eliminate the PMA of a film/substrate pair and alter anisotropy field. While these effects may arise unintentionally, one can also use these effects deliberately for engineering garnet films for devices. Therefore, the sensitivity of anisotropy properties of garnet thin films such as anisotropy field and effective anisotropy energy density needs to be evaluated with respect to changes in film strain and saturation magnetic moment.

Figure 5 shows the sensitivity of the effective magnetic anisotropy energy density to deviation of both strain and saturation magnetization, $M_s$ for five PMA film/substrate combinations: (a) HoIG/GGG, (b) YIG/YAG, (c) SmIG/SGGG, (d) HoIG/TGG, and (e) SmIG/NGG). The negative sign of effective magnetic anisotropy energy indicates PMA. The change of anisotropy energy from negative to positive indicates a transition from PMA to in-plane easy axis. In these plots, calculated anisotropy energies are presented for saturation moments and strains scanned from 60% to 140% of tabulated bulk garnet $M_s$ and of the strains of fully lattice-matched films on the substrates. The color scale indicates the anisotropy energy density in erg·cm$^{-3}$. Although magnetocrystalline anisotropy energies are negative for all of the thin film rare earth garnets considered here, these terms are negligible with respect to shape anisotropy ($K_1$(300 K) ~ -5% of $K_{shape}$). Therefore, magnetoelastic anisotropy term must be large enough to overcome shape anisotropy. A derivation of anisotropy energy as a function of $M_s$ and strain in equations (8)-(10) shows that the negative $\lambda_{111}$ values and negative strain states (compressive strain) for garnet films in Fig. 5(a)-(e) (HoIG, YIG, SmIG) enable these films to have PMA. To retain PMA state; $\lambda_{111}$ must be negative and large assuming elastic moduli and the Poisson's ratio are constant. The necessary condition for maintaining PMA is shown in equation 11.

$$K_{eff} = K_{indu} + K_{shape} + K_1 \qquad (8)$$

$$K_{eff} = -\frac{3}{2}\lambda_{111}\frac{Y}{1-v}\varepsilon_{||} + 2\pi M_s^2 + K_1 \qquad (9)$$

$$K_{eff} = \frac{3}{2}\lambda_{111}\frac{Y}{1-v}|\varepsilon_{||}| + 2\pi M_s^2 + K_1 \qquad (10)$$

$$K_{eff} < 0 \quad \text{if} \quad \left|\frac{3}{2}\lambda_{111}\frac{Y}{1-v}|\varepsilon_{||}| + K_1\right| > 2\pi M_s^2 \qquad (11)$$

Relaxing each fully strained and lattice-matched thin film towards unstrained state ($\varepsilon \rightarrow 0$ or moving from left to right on each plot in Fig. 5 causes the magnetoelastic anisotropy energy term to decrease in magnitude and gradually vanish. The total anisotropy energy decreases in intensity for decreasing strain and constant $M_s$. When $M_s$ increases, shape anisotropy term also increases and overcomes magnetoelastic anisotropy term. As a result, higher $M_s$ for relaxed films (i.e. relatively thick and iron-rich garnets) may lose PMA. Therefore, one needs to optimize the film stoichiometry and deposition process conditions, especially growth temperature, oxygen partial pressure and film thickness, to ensure that the films are strained and stoichiometric. Since strains are less than 1% in Fig. 5(a), 5(c)-(e), these samples are predicted to be experimentally more reproducible. For YIG on YAG, as shown in Fig. 5(b), the strains are around 3%, which may be challenging to reproduce. The cases presented in Fig. 5(a)-(e) are the only cases among 50 film/substrate pairs where reasonable changes in $M_s$ and strain may lead to complete loss of PMA. The rest of the cases have not been found as sensitive to strain and $M_s$ variability and those predicted to be PMA are estimated to have stable anisotropy. Effective anisotropy energy plots similar to Fig. 5(a)-(e) are presented in the supplementary figures for all 50 film/substrate pairs.

While PMA is a useful metric for garnet films, the effective anisotropy energy of the films should also not be too high ($<$ a few $10^5$ erg·cm$^{-3}$) otherwise the saturation fields for these films would reach or exceed 0.5 Tesla (5000 Oe). Supplementary figures present the calculated effective anisotropy energy and anisotropy field values for all 50 film/substrate pairs for changing strain and $M_s$ values. These figures indicate that one can span anisotropy fields of about 300 Oersteds up to 12.6 Tesla in PMA garnets. For practical integrated magnonic devices, the effective anisotropy energy should be large enough to have robust PMA although it should not be too high such that effective anisotropy fields (i.e. saturation fields) would still be small and feasible. Engineered strain and $M_s$ through controlled oxygen stoichiometry may help keep anisotropy field low while retaining PMA. In addition, according to the recently published paper on magnetic anisotropy of HoIG[50], the lattice matching in case of the thick samples becomes challenging to sustain, and the strain relaxes inside the film. Thus, the decrease in the anisotropy field is one consequences of the lower strain state, which is an advantage for magnonics or spin-orbit torque devices. Below a critical thickness, HoIG grown on GGG has PMA. However, as the film reaches this critical thickness, the 40% or more strain relaxation is expected and the easy axis becomes in-plane. So thinner films are preferred to be grown in integrated device applications.

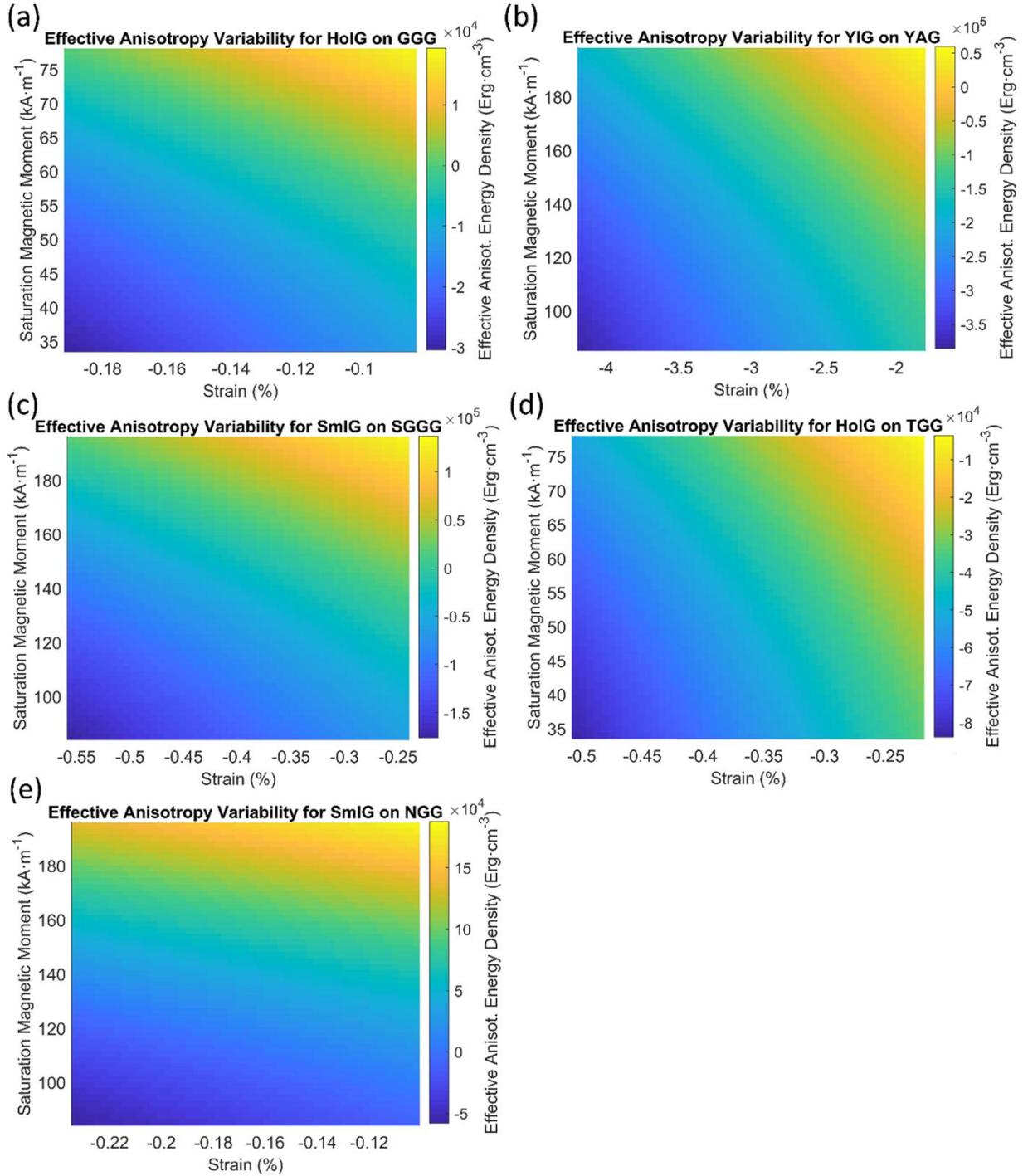

**Figure 5.** Effect of partial film relaxation and saturation magnetic moment variability on the effective anisotropy energy density of the films. Variation of effective magnetic anisotropy energy densities for (**a**) HoIG on GGG, (**b**) YIG on YAG, (**c**) SmIG on SGGG, (**d**) HoIG on TGG and (**e**) SmIG on NGG are presented when strain relaxation and magnetic saturation moments change independently. Film strain may vary from a completely lattice-matched state to the substrate to a relaxed state or a highly strained state due to microparticle nucleation. Strain variability alters magnetoelastic anisotropy and cause a PMA film become in-plane easy axis. On the other hand,

magnetic saturation moments may deviate from the tabulated values because of process-induced off-stoichiometry in the films (i.e. rare earth ion to iron ratio or iron deficiency or excess, oxygen deficiency). Relaxing the films reduces the magnetoelastic anisotropy term and diminishes PMA. Increasing $M_s$ strengthens shape anisotropy and eliminates PMA for low enough strains for all five cases presented.

Minimizing Gilbert damping coefficient in garnet thin films is also an important goal for spintronic device applications. First principles predictions of physical origins of Gilbert damping [51] indicate that magnetic materials with lower $M_s$ tend to have lower damping. Based on this prediction, DyIG, HoIG and GdIG films are predicted to have lower Gilbert damping with respect to the others. Since the compensation temperatures of these films could be engineered near room temperature, one may optimize their damping for wide bandwidths all the way up to terahertz (THz)[52] spin waves or magnons. The first principles predictions also indicate that higher magnetic susceptibility ($\chi m$) in the films helps reduce damping (i.e. lower saturation field). Therefore, the PMA garnet films with lower anisotropy fields are estimated to have lower Gilbert damping parameters with respect to PMA garnets with higher anisotropy fields.

**Conclusion**

Shape, magnetoelastic and magnetocrystalline magnetic anisotropy energy terms have been calculated for ten different garnet thin films epitaxially grown on five different garnet substrates. Negative $K_{eff}$ (effective magnetic anisotropy energy) corresponds to perpendicular magnetic anisotropy in the convention used here. By choosing a substrate with a lattice parameter smaller than that of the film, one can induce compressive strain in the films to the extent that one can always overcome shape anisotropy and achieve PMA for large and negative $\lambda_{111}$. Among the PMA films predicted, SmIG possesses a high anisotropy energy density and this film is estimated to be a robust PMA when grown on all five different substrates.

In order to obtain PMA, magnetoelastic anisotropy term must be large enough to overcome shape anisotropy. Magnetoelastic anisotropy overcomes shape anisotropy when the strain type (compressive or tensile) and magnetoelastic anisotropy constants $\lambda_{111}$ of the garnet film have the correct signs (not necessarily opposite or same) and the magnetoelastic anisotropy term has a magnitude larger than shape anisotropy. Both compressive and tensile-strained films can, in principle, become PMA as long as shape anisotropy can be overcome with large magnetoelastic strain effects. Here, in almost all cases that yield PMA on the given substrates, PMA iron garnets form under compressive lattice strain, except TbIG on SGGG and TbIG on NGG. These two cases have tensile strain and relatively large magnetocrystalline anisotropy, which could already overcome shape anisotropy without strain. Experiments are therefore suggested to target mainly compressive lattice strain.

20 different garnet film/substrate pairs have been predicted to exhibit PMA and their properties are listed on Table 2. For 7 of these 20 potential PMA cases, we could find unambiguous experimental demonstration of PMA. Among the 20 PMA cases, HoIG/GGG, YIG/YAG, SmIG/SGGG, HoIG/TGG and SmIG/NGG cases have been found to be sensitive to fabrication process or stoiochiometry-induced variations in $M_s$ and strain. In order to control effective anisotropy in rare earth iron garnets (RIGs), shape anisotropy could be tuned by doping garnet films with Ce [53], Tb [31] and Bi [54] or by micro/nano-patterning. Saturation magnetization could also be increased significantly by doping, which results in increasing the shape anisotropy in the magnetic thin films. Among the cases predicted to possess PMA, anisotropy fields ranging from 310 Oe (0.31 T) to 12.6 T have been calculated. Such a wide anisotropy field range could be spanned and engineered through strain state, stoichiometry as well as substrate choice. For integrated magnonic devices and circuits, garnets with low $M_s$ and lower anisotropy fields ($H_A < 0.5$ T) would require less energy for switching and would be more appropriate due to their lower estimated Gilbert damping.

**Methods**

**Calculation of anisotropy energy density.** We used $K_{eff} = K_{indu} + K_{shape} + K_1$ equation to calculate the total anisotropy energy density for each thin film rare earth iron garnet/substrate pair. Each anisotropy term consist of the following parameters: $K_{eff} = -\frac{3}{2}\lambda_{111}\frac{Y}{1-\nu}\varepsilon_{||} + 2\pi M_s^2 + K_1$. The energy density is calculated based on the parameters reported in previous references[16,34,36,40,41,43]. First order magnetocrystalline anisotropy, $K_1$, is an intrinsic, temperature-dependent constant reported for each REIG material. Young's modulus (Y), poison ratio (ν) and magnetostriction constant ($\lambda_{111}$) parameters evolving in the magnetoelastic anisotropy energy density term (first term) are considered to be constant according to the values previously reported. For shape anisotropy energy calculations (second term), bulk saturation magnetization ($M_s$) for each film was used. Since each film may exhibit variability in $M_s$ with respect to bulk, the model presented here yields the most accurate predictions when the actual film Ms, $\lambda_{111}$, Y, ν and $K_1$, and in-plane strain are entered for each term. The original Microsoft Excel and MATLAB files used for generating the data for Figures 1-5 are presented in the supplementary files.

## Acknowledgments

M.C.O. acknowledges BAGEP 2017 Award and TUBITAK Grant No. 117F416.

## Competing interests

There is no financial and non-financial competing interest among the authors.

## Author contributions

M.C.O. designed the study. S.M.Z. performed the calculations and evaluated and analyzed the results with M.C.O. Both authors discussed the results and wrote the manuscript together.